\documentclass[conference]{IEEEtran}
\usepackage{blindtext, graphicx, tabularx, booktabs, amsmath, siunitx, subfigure, xcolor}

\usepackage{cite}

\usepackage{color}
\usepackage{soul}
\usepackage{flushend}
\usepackage{pdfpages}

\newcolumntype{R}[1]{>{\raggedright\arraybackslash}p{#1}}

\IEEEoverridecommandlockouts

\begin{document}

\begin{titlepage}
\includepdf{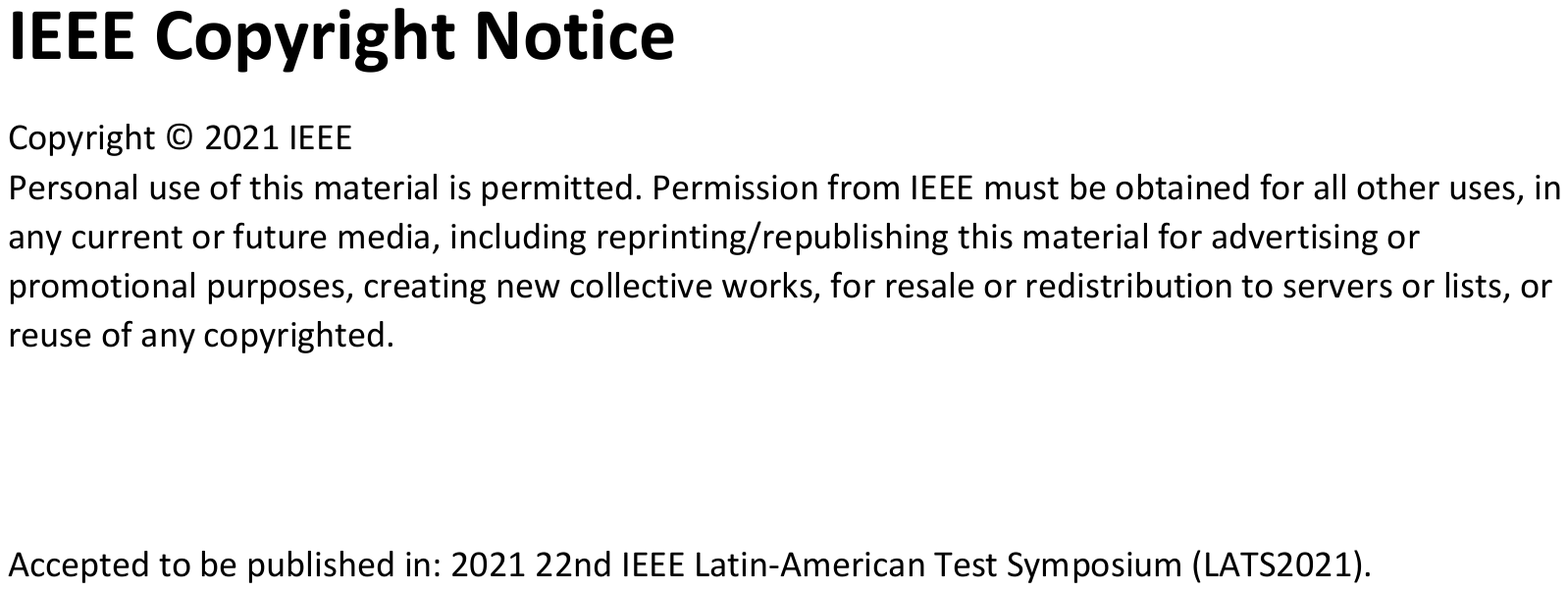}
\end{titlepage}

%
\title{

A Tutorial on Design Obfuscation: \\from Transistors to Systems


\thanks{

This work has been partially conducted in the project ``ICT programme'' which was supported by the European Union through the European Social Fund. It was also partially supported by the Estonian Research Council grant MOBERC35.}

}

 \author{
     \IEEEauthorblockN{
         Samuel Pagliarini
     }
  
      \IEEEauthorblockA{
          Tallinn University of Technology (TalTech) \\ Department of Computer Systems \\ Centre for Hardware Security
          \\E-mail: samuel.pagliarini@taltech.ee
      }
}

\maketitle

\begin{abstract}
The recent advances in the area of design obfuscation are encouraging, but may present themselves as hard to read for a non-specialist audience. This tutorial uncovers these advances in a clear language, contrasting the approaches that can be implemented at layout level, in the netlist of a circuit, or even at chip level. This tutorial also highlights the available support, both from the tooling side and the logistics of fabricating an obfuscated integrated circuit.

\end{abstract}
    

\section{Introduction}
\label{sec_introduction}

Historically, we have seen a constant offshoring trend for integrated circuit (IC) foundries that is supported by the fabless business model that most design companies subscribe to. Eventually, this trend that has been going on for decades, started to raise alarms that governments would lose access to state-of-the-art silicon \cite{Lieberman2003} and instead rely on offshore foundries. That being said, foundries are typically considered a potential adversary in this scenario since the design house that creates the IC is not a stakeholder of the foundry. Compounded by geographical and political concerns, these foundries are then termed as \emph{untrusted foundries}. 

When it comes to attacks that can be mounted by untrusted foundries, the ICs and their many intellectual property (IP) modules are vulnerable to a diverse selection of threats. We direct the reader to \cite{Rostami2014} for a thorough discussion on the supply chain and related threats, while here we provide a short description of commonly studied threats:
\begin{itemize}
\item \textbf{Hardware trojans}. Malicious modification or insertion of logic into an IC.
\item \textbf{IP piracy}. Unauthorized copy of an IP, by the foundry or any other user of the IP.
\item \textbf{IC overbuilding} Production of parts beyond the contracted amount with the goal of reselling. 
\item \textbf{Reverse engineering} Process of deconstruction, either physical or logical, of an IP or IC with the intent of extracting knowledge from it.
\end{itemize}

Obfuscation has emerged as an interesting solution for directly combating IP piracy and IC overbuilding, while also presenting some implications for the other threats. Let us now discuss some of the more relevant obfuscation techniques in the specialized literature.

\section{Current Practices in Obfuscation}
\label{sec_related}

\subsection{Layout Obfuscation}

In \textbf{camouflaged logic}, the central concept is to make one logic gate look like another, such that an attacker cannot discern the functionality of a particular gate based solely on its observable physical characteristics. One proposed camouflaging technique \cite{Rajendran2013b} uses a mix of dummy and real vias/contacts to obscure the gate's function. Dummy vias/contacts are not fully formed and do not make an electrical connection between layers, but they possibly may be mistaken for real vias/contacts during the delayering process of reverse engineering a fabricated circuit. The proposed design in \cite{Rajendran2013b} uses a camogate that can form a NAND, NOR, or XOR gate with identical layouts (except for the placement of real and dummy vias). The security of this technique relies on the limited ability of the attacker to distinguish between real and dummy vias/contacts.

\begin{figure}
	\centering
	\includegraphics[width=0.75\linewidth]{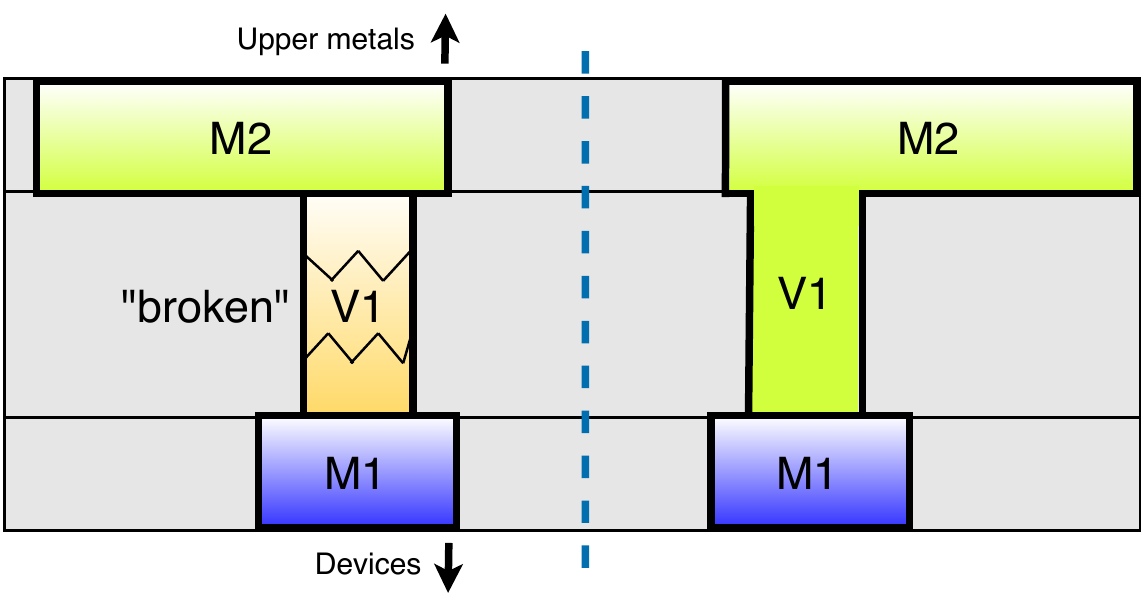}
	\caption{Via and metal deposition processes. On the left, conventional deposition where vias and metals are formed one by one. On the right, the dual damascene process is highlighted where metals and vias are deposited at the same time.}
	\label{fig:via}
\end{figure}

However, the delayering and imaging capabilities of state-of-the-art reverse engineering labs are very advanced \cite{Chipworks} and may not be fooled by such dummy vias/contacts. Further, dummy via/contact fabrication is caught in a dual-sided constraint between reliability and security as illustrated in Fig. \ref{fig:via}. If the dummy has a large separation between layers, then reliability is not compromised, but it becomes more easily distinguished as a dummy. Conversely, a small separation is harder to distinguish, but is more likely to create a short. Additionally, the compatibility of building dummy `broken' vias in a modern IC is questionable, as the fabrication process forms the vias in the same manufacturing step as the interconnect wires themselves. It is unclear how such a partially connected via/contact could be formed without additional special process steps and masks, increasing the cost and complexity of implementing this countermeasure. The contrast between conventional interconnect deposition and present deposition is shown in Fig. \ref{fig:via}. 

\begin{figure}
    \centering
	\includegraphics[width=0.68\linewidth]{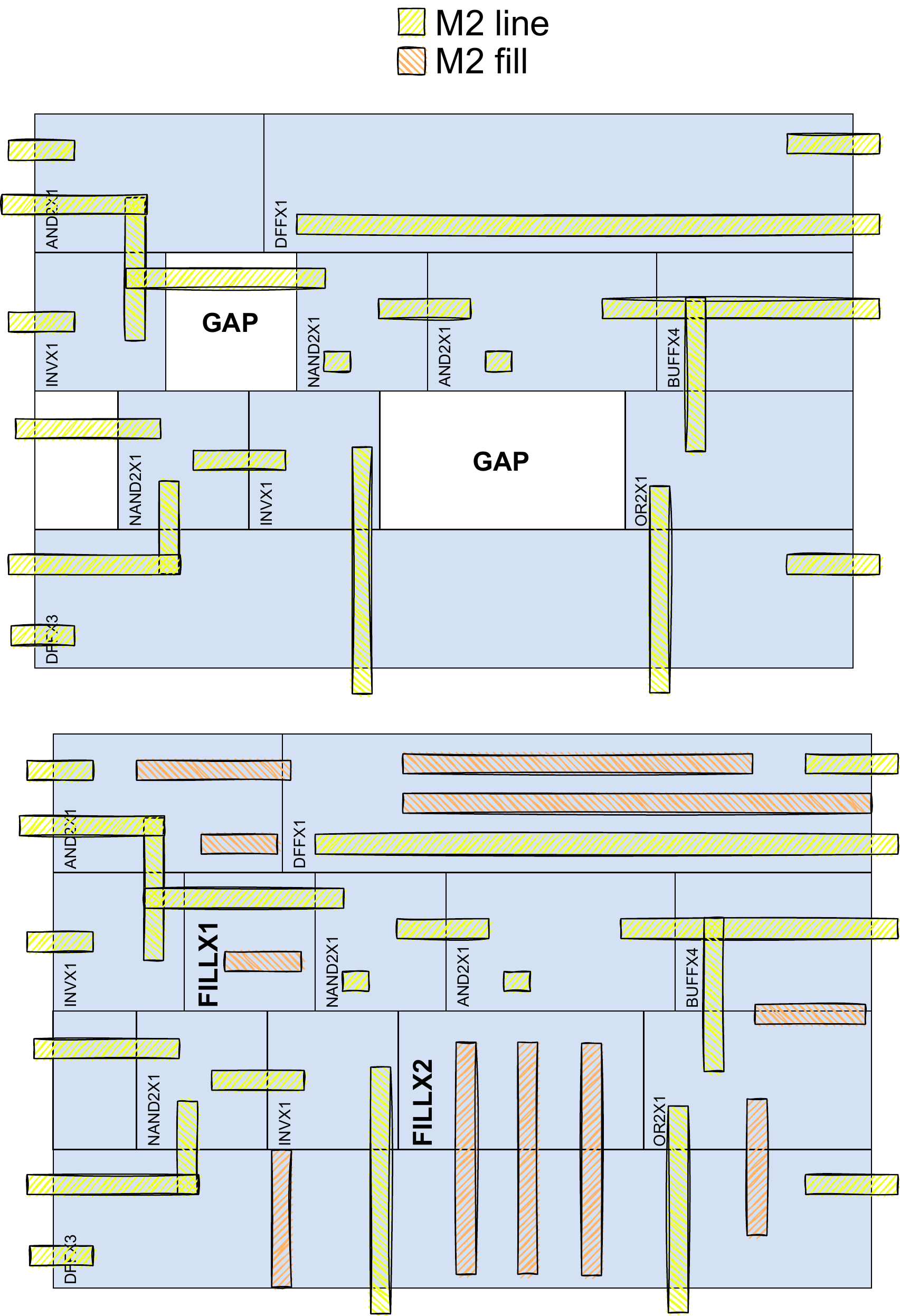}
	\caption{Layout of a placed standard cell-based circuit, before and after fill procedures. The fill lines are overly simplified.}
	\label{fig:fill}
\end{figure}

Another class of layout obfuscation techniques is worth mentioning: layout manipulation during place \& route to avoid ‘white spaces’ \cite{Hossein2017}. But this is partially misleading: empty standard cell rows/slots eventually get filled by filler cells, while unused routing tracks may get filled with filler wires for DFM reasons. This effect is shown in Fig. \ref{fig:fill} where we highlight only M2 for clarity. Notice how FILLX cells are placed in the gaps between standard cells and how specialized M2 fill lines are positioned in areas of low density.

The design rules for today’s advanced nodes forcibly imply critical layers are gridded and regular, including poly, active, and fin layers. Effectively, circuits being fabricated today already resemble a sea of gates at their lowest layers, leaving even less margin for layout manipulation for obfuscation. Overall, layout modifications are \emph{very limited} in combating threats coming from an untrusted foundry. Tooling support is largely non-existent and \emph{ad hoc} solutions are employed.

\subsection{Key-based Locking}

\begin{figure}
	\centering
	\includegraphics[width=0.50\linewidth]{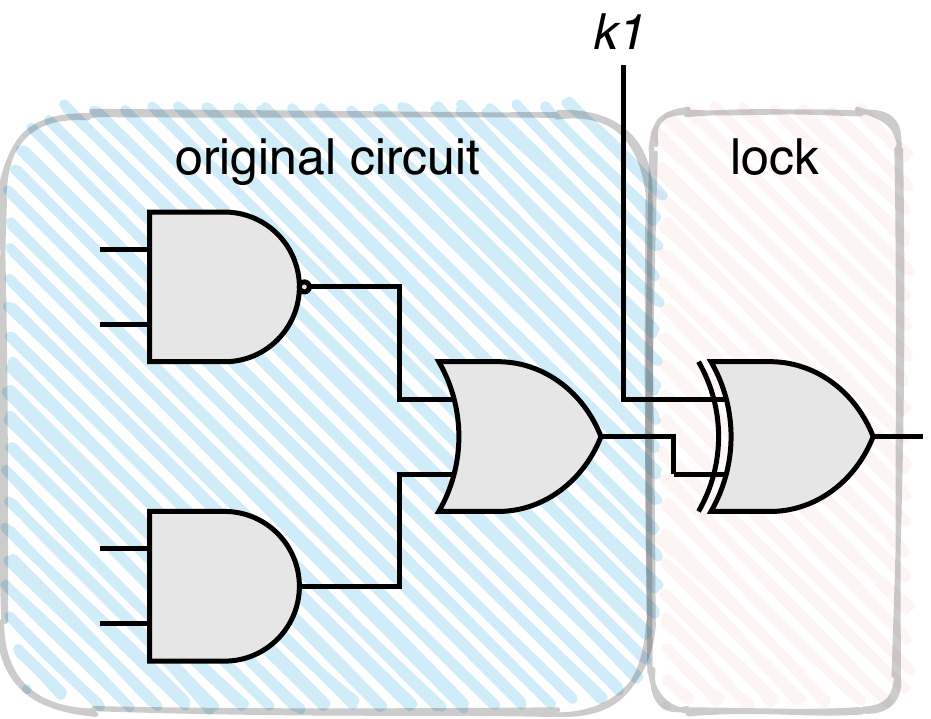}
	\caption{Example of a locked circuit: the lock is the XOR gate controlled by key \emph{k1}. The circuit behaves as expected when \emph{k1=0}.}
	\label{fig:lock}
\end{figure}

Designs obfuscated with key-based locking require a key to operate correctly. The security strength of the technique comes from the fact that the key is not shared with an untrusted foundry. \textbf{Logic locking} is a class of techniques that has evolved over the last decade or so and gained significant traction. Some authors use other terms to refer to this class of techniques, such as logic encryption, logic obfuscation, or keyed logic. Conventionally, logic locking \cite{Roy2008} has been considered for preventing reverse engineering, excess production, and, to a lesser extent, to prevent malicious circuit modifications. The original concept is relatively simple: additional logic is embedded into a circuit to provide it with a key-based operation. With the correct key, the circuit operates as intended. Without the correct key, some selected signals are corrupted. The key-based operation is usually implemented with XOR/XNOR gates, but MUXes and other logic cells can also be used for the same effect. A minimal example of an XOR-based lock is given in Fig. \ref{fig:lock}. For a large circuit, a number of XORs and keys would then be used, given that the difficulty to defeat logic locking increases with the number of locks.

\begin{figure}
	\centering
	\includegraphics[width=0.95\linewidth]{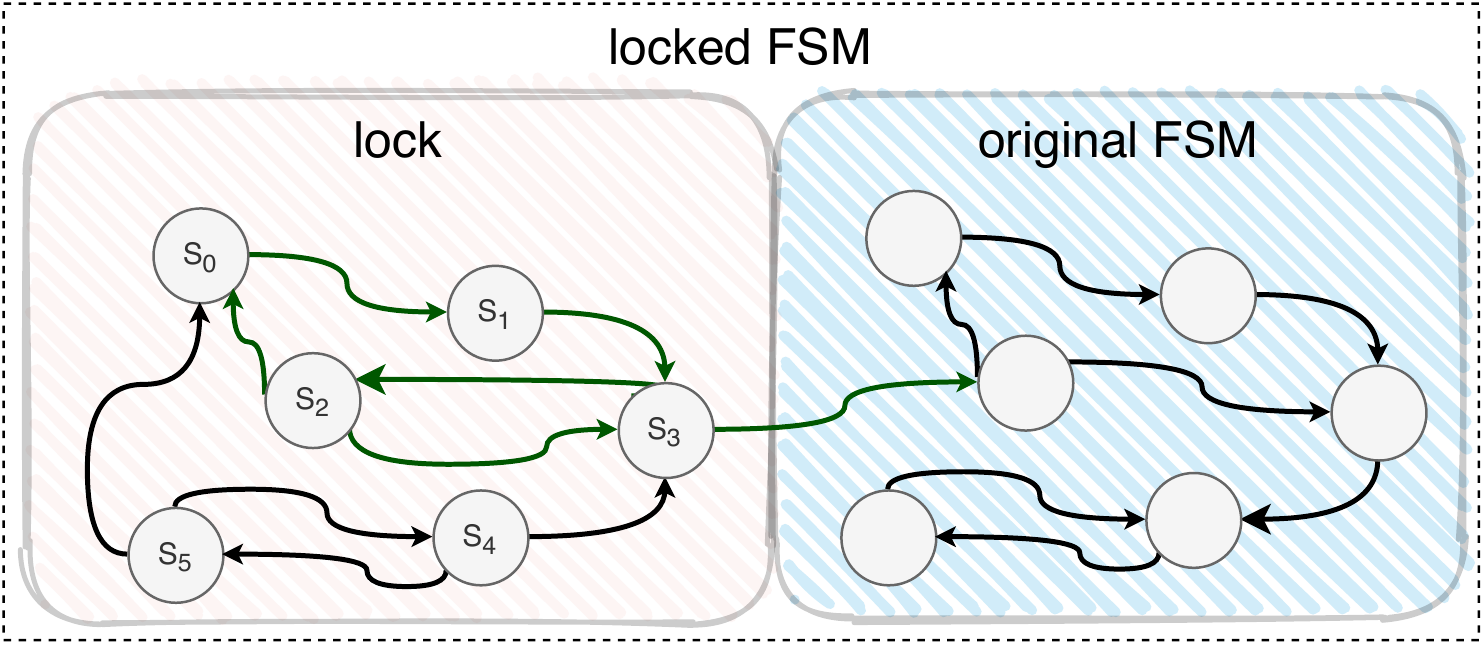}
	\caption{FSM locking scheme where the transition $\text{S}_0\text{-S}_1\text{-S}_3\text{-S}_2\text{-S}_3$ has to be exercised to reach the original FSM.}
	\label{fig:fsm}
\end{figure}

\textbf{FSM locking} is the sequential counterpart of logic locking. Instead of additional combinational gates, new states and transitions are added to an existing FSM in order to create a locked FSM \cite{Chakraborty2009}. When the right key is applied, the locked FSM transitions from state to state until the original FSM becomes accessible and from there onwards, the circuit behaves as the original one would. This scheme is shown in Fig. \ref{fig:fsm}. The cost of implementing FSM locking is relatively small: a few flops for new states and some combinational cells for new transitions, which can have close to no impact on the power, area, or performance of a system.

 Overall, the effectiveness of key-based approaches is being constantly challenged \cite{subramanyan15}. However, tooling support is vast \cite{cadforassurance}.
 
 \begin{figure*}[t!]
	\centering
	\includegraphics[width=0.9\linewidth]{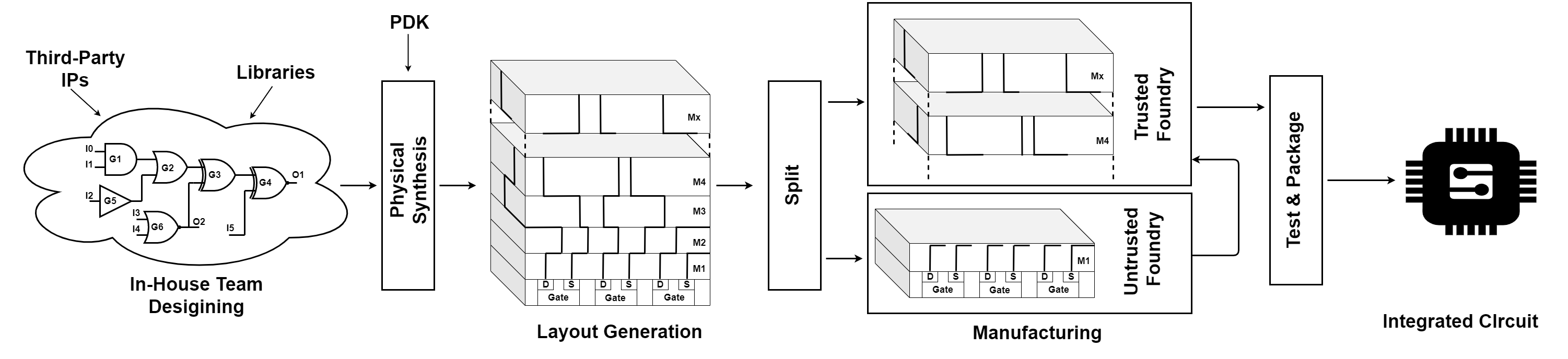}
	\caption{Split Manufacturing Design Flow \cite{split}.}
	\label{fig:split}
\end{figure*}

\subsection{Macro approaches}
In this obfuscation class, approaches make use of (some degree of) trusted fabrication, i.e., these macro approaches rely on foundry collaboration/trust that cannot be achieved by design means only. \textbf{Split fabrication} \cite{split} is the most well-known example in this category. As illustrated in Fig. \ref{fig:split}, a split-manufactured design is cut horizontally and different foundries are in charge of fabricating the top and bottom parts. Here, the typical assumption is that an untrusted foundry will mostly handle the fabrication of device layers while a trusted foundry will handle the interconnect. By doing so, the untrusted foundry does not fully understand how the transistors are connected, which by itself mitigates the piracy and overbuilding threats. The scheme is also reportedly efficient against hardware trojans.

Split fabrication suffers from one enormous drawback: the typical foundry business model does not contemplate the delivery of `unfinished' wafers. On the contrary, foundries deliver wafers (or cut dice) that contain the entire stack, from devices to the uppermost layer metal (e.g., M10). In order to bypass this limitation and still benefit from the state-of-the-art transistors that untrusted foundries offer, a \textbf{Split-Chip} technique is defined in \cite{dandt}. The concept is to create a vertical cut in the design, where modules are partitioned and assigned to either a trusted or untrusted side. If properly split, the two designs can be bound together with minimal performance and power penalties. Generally speaking, modules that are more control-oriented are good candidates for the trusted side of the system while computation-intensive modules are good candidates for the untrusted side.

Unfortunately, despite the obvious advantages, macro solutions require foundry cooperation and no clear framework exists for this activity today. Tools for implementing the aforementioned macro solutions are typically implemented as scripts within commercial CAD tools.

\section{Discussion and conclusions}

Other active lines of research in obfuscation, although not covered in this tutorial, are the use of CMOS augmented by other materials (e.g., non-volatile storage of the circuit state with magnetic elements \cite{magnetic}) or paradigms (e.g., selective use of eFPGA\cite{e-FPGA2}). However, these can still be considered emerging approaches. 

While the literature offers a wide range of solutions for obfuscation, it is not possible to indicate a clear one-size-fits-all technique. From an academic perspective, logic locking is by far the most popular approach and an incredibly active research topic. While hardware security remains an afterthought in many instances, logic locking has the potential to break that barrier: It is conceivable that logic locking will be considered for protecting mass produced devices (i.e., consumer electronics) in the near future. For this reason, practitioners should familiarize themselves with the technique and its variants.

\bibliographystyle{IEEEtran}
\bibliography{main}

\end{document}